\documentclass[%
 aip,
 amsmath,amssymb,
 reprint,%
]{revtex4-1}

\usepackage{graphicx}
\usepackage{dcolumn}
\usepackage{bm}
\usepackage{float}
\usepackage{siunitx}
\usepackage[utf8]{inputenc}
\DeclareUnicodeCharacter{2212}{\textminus}
\usepackage[T1]{fontenc}
\usepackage{mathptmx}
\usepackage{etoolbox}
\usepackage{gensymb}
\usepackage{upgreek}

\makeatletter
\def\@email#1#2{%
 \endgroup
 \patchcmd{\titleblock@produce}
  {\frontmatter@RRAPformat}
  {\frontmatter@RRAPformat{\produce@RRAP{*#1\href{mailto:#2}{#2}}}\frontmatter@RRAPformat}
  {}{}
}%
\makeatother

\newcommand{\vb}{$V_{\rm B}^-$}
\newcommand{\vbn}{$V_{\rm B}$}
\begin{document}

\title{Impact of irradiation conditions on the magnetic field sensitivity of spin defects in hBN nano flakes}
\author{Saksham Mahajan}
\affiliation{Department of Electrical and Electronic Engineering, UCL, London WC1E 7JE, UK.}
\email{saksham.mahajan@ucl.ac.uk}

\author{Ravi Kumar}%
\affiliation{London Centre for Nanotechnology, UCL, 17-19 Gordon St, London WC1H 0AH, UK.}

\author{Aferdita Xhameni}%
\affiliation{Department of Electrical and Electronic Engineering, UCL, London WC1E 7JE, UK.}

\author{Gautham Venu}
\affiliation{Department of Electrical and Electronic Engineering, UCL, London WC1E 7JE, UK.}

\author{Basanta Mistri}
\altaffiliation[Also at ]{Center for Quantum Information, Communication and Computing, Indian Institute of Technology Madras, Chennai 600036, India}
\affiliation{Department of Physics, Indian Institute of Technology Madras, Chennai 600036, India}

\author{Felix Donaldson}
\affiliation{London Centre for Nanotechnology, UCL, 17-19 Gordon St, London WC1H 0AH, UK.}


\author{T. Taniguchi}
\affiliation{Research Center for Materials Nanoarchitectonics, National Institute for Materials Science,  1-1 Namiki, Tsukuba 305-0044, Japan}

\author{K. Watanabe}
\affiliation{Research Center for Electronic and Optical Materials, National Institute for Materials Science, 1-1 Namiki, Tsukuba 305-0044, Japan}

\author{Siddharth Dhomkar}
\altaffiliation[Also at ]{Center for Quantum Information, Communication and Computing, Indian Institute of Technology Madras, Chennai 600036, India}
\affiliation{London Centre for Nanotechnology, UCL, 17-19 Gordon St, London WC1H 0AH, UK.}
\affiliation{Department of Physics, Indian Institute of Technology Madras, Chennai 600036, India}

\author{Antonio Lombardo}
\affiliation{Department of Electrical and Electronic Engineering, UCL, London WC1E 7JE, UK.}
\affiliation{London Centre for Nanotechnology, UCL, 17-19 Gordon St, London WC1H 0AH, UK.}

\author{John J.L. Morton}
\affiliation{Department of Electrical and Electronic Engineering, UCL, London WC1E 7JE, UK.}
\affiliation{London Centre for Nanotechnology, UCL, 17-19 Gordon St, London WC1H 0AH, UK.}

\date{\today}

\begin{abstract}
We study \vb\ centres generated by helium focused ion beam (FIB) irradiation in thin ($\sim$70~nm) hBN nanoflakes, in order to investigate the effect of implantation conditions on the key parameters that influence the magnetic field sensitivity of \vb\ quantum sensors. 
Using a combination of photoluminescence, optically detected magnetic resonance, and Raman spectroscopy we examine the competing factors of maximising signal intensity through larger \vb\ concentration against the degradation in spin coherence and lattice quality observed at high ion fluences.
Our results indicate that both the \vb\ spin properties and hBN lattice parameters are largely preserved up to an ion fluence of 10$^{14}$ ions/cm$^2$, and beyond this significant degradation occurs in both. At the optimal implantation dose, an AC magnetic sensitivity of $\sim 1 \upmu\text{T}/\sqrt{\text{Hz}}$ is achieved. Using the patterned implantation enabled by the FIB we find that \vb\ centres and the associated lattice damage are well localized to the implanted regions.
This work demonstrates how careful selection of fabrication parameters can be used to optimise the properties of \vb\ centres in hBN, supporting their application as quantum sensors based in 2D materials.
\end{abstract}

\maketitle

\section{\label{sec:level1}Introduction\protect\\}
Optically active point defects in solid-state materials have long attracted interest for the development of quantum technologies\cite{quantuminformation,quantumsesningreviewcdegen,Taylor2008,morello2010single,de2008nmr,magneticsensingvalueGottscholl2021,biosensing} due to their long quantum coherence times, simple measurement protocols and scalability.  For instance, point defects in three-dimensional materials, such as color centres in diamond\cite{nvcenter},  have been widely explored for nanoscale quantum sensing with applications ranging from probing current distributions in materials such as graphene\cite{graphene} to in-situ studies of biological systems, including single proteins\cite{protein} and living cells\cite{cell}. Although such point defects offer lower absolute magnetic field sensitivity than state of the art sensors like SQUIDs ($\eta_{mag} \sim$ 1~fT/$\sqrt{Hz}$ )\cite{nvsensing}, their atomistic nature and presence within wide band gap materials enables higher spatial resolution, and robust operation across a broad temperature range. To take advantage of such enhanced spatial resolution, these defects must be positioned close to the surface, where interactions with surface-related noise sources degrade quantum coherence and ultimately limit sensitivity.

As a potential solution to the limitations of surface noise on the performance of quantum sensors in three-dimensional materials like diamond, wide bandgap van der Waals (2D) materials have been explored as hosts for addressable quantum systems. 
Among these, the negatively charged boron vacancy (\vb) centre in hexagonal boron nitride (hBN)  has been identified as an optically addressable spin\cite{boronvacancyesrGottscholl2020, magneticsensingvalueGottscholl2021} with the ability to sense various external physical stimuli such as electric and magnetic fields, temperature and pressure \cite{Gottscholl2021SpinSensors}. Despite their short coherence time (Hahn echo $T_2$ < 100~ns)\cite{isotopeengineeringandt2matchingusHaykal2022,coherentspindynamicsnatureGong2022} arising from the hBN nuclear spin bath, and their low quantum efficiency ($\sim$ 0.03\%)\cite{lowefficieny}, the prospect of being able to position optically addressable spins in 2D materials with sub-nanometer proximity to the target sample offers potential advantages for sensing applications.

Boron-vacancy centres can be created using ion implantation, which offers advantages such as spatial localization through use of a focused ion beam (FIB) and improved repeatability compared to alternative methods\cite{guo_generation_2022}. Prior studies have successfully demonstrated the creation of \vb\ centres using a variety of ions (H, He, Ga, N)~\cite{fibcite, Kianinia2020,raman2} under different implantation conditions. 
These implantation parameters must be carefully optimized to achieve a high concentration of \vb\ centres while maintaining the hBN crystallinity. Although a higher concentration of \vb\ centres enhances signal intensity which, in principle, improves magnetic field sensitivity, increasing the implantation fluence can affect spin coherence time and introduce lattice damage, which hinders the fabrication of hBN based heterostructures.

In this study, we employ helium FIB at various doses to form \vb\ centres in hBN and then study their spin and optical properties. We evaluate how varying ion fluence alters factors such as count rate, coherence, optical contrast, and the extent of lattice damage, which ultimately contribute to the \vb\ magnetic field sensitivity. FIB implantation was performed using low energy He\(^+\) ions (25~keV) across a range of ion fluences ($10^{12}$ to $10^{15}$~ions/cm$^2$) in hBN nanoflakes with thickness $\sim 70$~nm. Helium ions were preferred due to their low atomic mass, which minimizes lattice damage\cite{multipledefectshe} and their ability to introduce a high density of point defects\cite{multipledefectshe}. 
Photoluminescence (PL), spin resonance, and Raman spectroscopy reveal a correlation between reduced coherence times and increased lattice disorder. The optimum AC magnetic field sensitivity (approximately 1~$\upmu\mathrm{T}/\sqrt{\mathrm{Hz}}$) was achieved at an ion fluence of $10^{14}$~ions/cm$^2$, where hBN crystallinity is preserved. Finally, we analyze the spatial distribution of defects around the patterned implantation interface using PL and Raman spectroscopy. 

\begin{figure}[t] 
    \centering
    \includegraphics[width=0.48\textwidth]{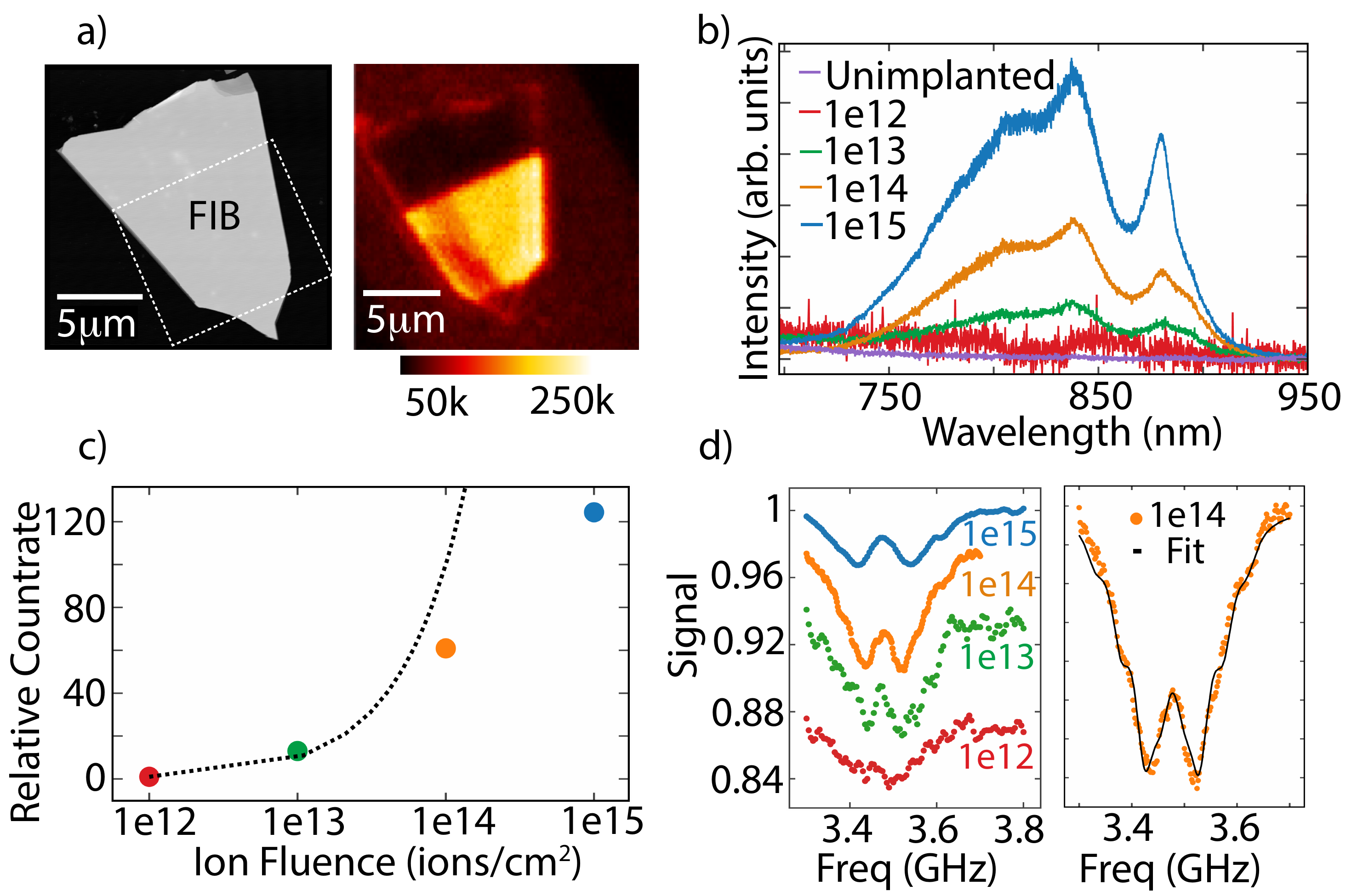} 
    \caption{ Fabrication and optical characterization of \vb\ centres in hBN nanoflakes. a) AFM (left) and fluorescence confocal image (right) of a sample irradiated with an ion fluence of $10^{14}$~cm$^{-2}$. The interfaces between implanted and unimplanted regions are visible in the confocal image. b) PL spectra of unimplanted and implanted samples, with the latter showing characteristic \vb\ emissions centered around $\sim 820$~nm. To aid comparison of spectral features, the PL spectra of samples 1e12 and 1e13 have been scaled in amplitude by a factor of 10 and 2, respectively. c) Relative fluorescence intensity (calculated from PL spectra) of different samples with dotted line representing linear increase in counts with ion fluence. d) Pulsed ODMR spectra (left) along with fitted data for sample 1e14 (right) using the theoretical model (for details, see SI). } 
    \label{fig:wide-image}
\end{figure}

\section{Experimental Details\label{sec:level2}}

\subsection{Sample fabrication}

The hBN samples were prepared by mechanical exfoliation of bulk crystals which were grown under high pressure and temperature \cite{taniguchi2007synthesis}. Subsequently, four hBN nanoflakes with similar thicknesses ($\sim 70$~nm) were identified through atomic force microscopy (Bruker Dimension Icon with ScanAsyst) and an example image is shown in Fig.~\ref{fig:wide-image}(a).
The selected hBN nanoflakes were transferred onto a gold microstrip (thickness 200~nm, width 20~$\upmu$m and length 3.5~mm, fabricated on a silicon substrate)
for plasmonic enhancement of \vb\ related emission \cite{plasmonicenhancementGao2021} as well as efficient microwave delivery.
After the flake transfer, \vb\ centres were created using FIB (ZEISS ORION NanoFab) implantation of He$^+$ ions (25~keV) with fluences ranging between $10^{12}$ and $10^{15}$~{ions/cm}$^2$. Defect distributions were estimated using Stopping and Range of Ions in Matter (SRIM) calculations (Fig.~SI2) with SRIM 2008 software, indicating nearly uniform \vbn\ defect concentration along the flake thickness. 
To enable characterization of the defect distribution around the implantation interface, a portion of the nanoflake was left unimplanted, as illustrated in the confocal image of Fig.~\ref{fig:wide-image}(a). Below, we refer to the four different samples by the ion fluence used, in ions/cm$^2$. 

\subsection{Methods}

Photoluminescence and optically detected spin measurements were performed using a home-built confocal microscopy setup (NA = 0.65)  equipped with a Montana Instruments s100 cryostation and a 522~nm continuous-wave laser (LBX-522; Oxxius). The excitation laser power on different samples varied between 4 and 6~mW. The PL signal was filtered using either a 550~nm or 750~nm long-pass filter (LPF), depending on the experimental requirement, and guided toward either the single-photon counting module (Excelitas Technologies) or the PL spectrometer (SpectraPro HRS500, Princeton Instruments). 
To investigate the spin properties of the \vb\ centres, pulsed optically detected magnetic resonance (PODMR) spectroscopy was performed under zero applied  magnetic field. 
For microwave excitation, a Rohde \& Schwarz SGS100A source was modulated through a Swabian Pulse Streamer 8/2. The microwave signal was amplified using a Mini-Circuits amplifier (ZVE-3W-83 +) and delivered to the sample via a gold microstrip wire-bonded to a co-planar waveguide (see Fig.~SI3). PODMR spectra were digitally filtered using 35~MHz moving average and 
fitted using a spin Hamiltonian model (implemented using QuTiP) which incorporates a single boron vacancy coupled to the three first-shell nitrogen atoms \cite{theoreticalmodel1,theoreticalmodel2}. 
For PODMR fitting, the transition probabilities for all possible spin transitions were calculated using Fermi's Golden Rule. 
Additional details about the theoretical model are provided in the SI.

Spin-lattice relaxation times ($T_1$) of \vb\ spins were measured by varying the delay between initialization and readout laser pulses, and spin coherence times ($T_2$) using a Hahn echo sequence, with phase cycling of the initial pulse to suppress free induction decay (FID) signals superimposed on the decay\cite{fidremoval}. 
Raman spectra were obtained using a Bruker Senterra II Raman spectrometer system 
and recorded between 1000--2000~cm$^{-1}$ using 532~nm laser excitation ($\sim$2.5~mW of laser power focused through a 0.6~NA objective, resulting in a spot size < 1~$\upmu$m). 
%
Additionally, to study the point defect-related features ($\sim$1290~cm$^{-1}$) in sample 1e15, multi-wavelength Raman spectroscopy was performed between laser wavelengths of 457--785~nm. For spectral measurements at 457~nm and 514~nm, a separate spectrometer system (Renishaw inVia Raman Microscope, N.A. 0.75, laser power $\sim$ 2~mW) was used.

\begin{figure}[htbp] 
    \centering
    \includegraphics[width=0.49\textwidth]{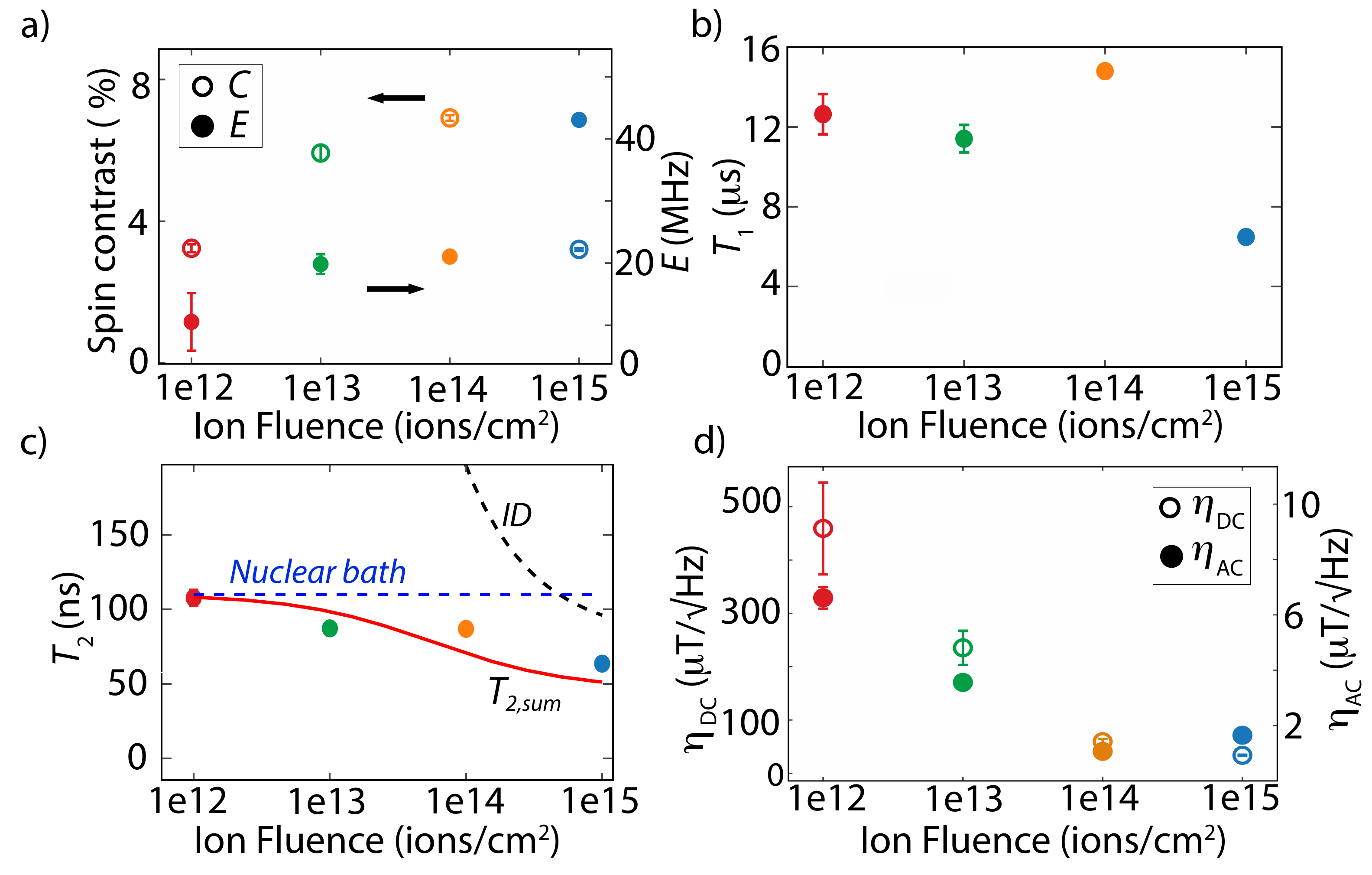} 
    \caption{Spin properties of \vb\ centres as a function of ion fluence. a) PODMR contrast (solid circles) and transverse strain splitting parameter ($E$, open circles). b) Spin-lattice relaxation time, T$_{1}$. (c) The decoherence time T$_{2}$ measured by Hahn echo, including various limiting factors: a decoherence rate from the nuclear spin bath (independent of ion fluence), and contributions from instantaneous diffusion (ID) calculated using the model described in the text. d) Estimated AC and DC magnetic sensitivity for the different samples.}
    \label{fig: spin}
\end{figure}

\section{\label{Results}Results and Discussion}

Photoluminescence spectroscopy (PL) of the FIB implanted flakes (see Fig.~\ref{fig:wide-image}(b)) reveal the characteristic \vb\ related sideband emission with a peak intensity at around 820~nm~$^[$\cite{boronvacancyesrGottscholl2020}$^]$. To investigate dependence of PL signal intensity on ion fluence, each spectrum was integrated over the wavelength range 750--950~nm, and the integrals were normalized to that of the 1e12 sample. The results, shown in Fig.~\ref{fig:wide-image} (c), show a sublinear increase in  \vb\ related PL emission with increasing ion fluence \cite{coherentspindynamicsnatureGong2022}, indicating a reduced conversion efficiency of ions to \vb\ centres at fluences above $10^{13}$~{ions/cm}$^2$. 

Pulsed ODMR measurements on each sample exhibit the characteristic resonance of \vb\ centres (around 3.5 GHz), as shown in Fig.~\ref{fig:wide-image}(d)). 
Fits to the PODMR spectra were used to extract spin Hamiltonian parameters $D$ and $E$, corresponding to the zero field splitting and strain splitting respectively. While the extracted $D$ was found to be largely independent of ion fluence (see Fig.~SI4), the strain splitting $E$ increases monotonically with increasing ion fluence (see Fig.~\ref{fig: spin}(a)). In particular, a significant increase in strain was observed for a fluence of $10^{15}$~{ions/cm}$^2$, indicating a threshold for the formation of extended defects. These extended defects introduced substantial lattice distortion, as confirmed by Raman spectroscopy (discussed further below in Fig.~\ref{fig:raman}. The observed variation in $E$ is consistent with previous reports in which different ion implantation parameters were compared in terms of their effects on spin properties of the \vb\ centres \cite{electronirradiationoptimization,coherentspindynamicsnatureGong2022,guo_generation_2022, ODMRbroadening}. 
A further impact of the lattice damage caused at higher ion fluences is seen in the PODMR signal contrast (also shown in (see Fig.~\ref{fig: spin}(a)). 
The signal contrast initially increases with increasing ion fluence (up to $10^{14}$~{ions/cm}$^2$) as the \vb\ emission becomes more prominent over the background. However, at the highest fluence of $10^{15}$~{ions/cm}$^2$, the increased lattice distortion 
mixes the spin states and reduces the ODMR contrast\cite{spinmixing}.
We investigated the spin-lattice relaxation times ($T_{1}$) of the different samples (see Fig.~\ref{fig: spin}(b)) and found no significant variation in $T_{1}$ up to a fluence of $10^{14}$~{ions/cm}$^2$. However, $T_{1}$ was significantly reduced (halving to $\sim$ 6~$\upmu$s) in sample 1e15. This behavior may also be attributed to increased lattice disorder, in this case whereby a resulting anharmonicity in the local phonon density of states enhances phonon-mediated relaxation processes~\cite{coherentspindynamicsnatureGong2022,decreasingt1doseoptimizationincreasingeGuo2022}.

The performance of \vb\ based quantum sensors depends more critically on the spin coherence time, $T_2$ as these are significantly shorter than $T_1$. We measured $T_2$ using a Hahn echo sequence, which decouples the \vb\ centres from the static components of the Ising interaction with surrounding nuclear spins. 
The resulting decays were well fit by a single exponential decay and the measured $T_2$ was found to gently  
decrease  
with increasing ion fluence across the range of samples studied (see Fig.~\ref{fig: spin}(c)). $T_2$ in the low-fluence limit (1e12) was found to be 100~ns, consistent with a limit from the nuclear spin bath, as established in previous studies \cite{coherentspindynamicsnatureGong2022}. As the fluence increases, additional decoherence mechanisms may arise from the increasing \vb\ spin concentration and the creation of other spin-active defects.
We first consider spectral diffusion experienced by the central spin arising from the spin flips of its neighbors (we neglect any potential contributions that might arise from spin flip-flops due to large inhomogeneous spin linewidth compared to average dipolar coupling strength $\sim 2$~MHz) \cite{id}. A quantitative estimate of the spectral diffusion limited T$_2$ yielded coherence times between 18~$\upmu$s (for 1e12) to 1.1~$\upmu$s (for 1e15 sample), based on the measured values of $T_1$, so its contribution can be neglected. We next consider so-called `instantaneous diffusion' which arises from dipolar coupling with \emph{resonant} spins which are also flipped by the Hahn echo $\pi$ pulse, such that the interactions are not refocused when the echo is formed. The decoherence rate due to instantaneous diffusion can be estimated using the following relation\cite {id,schweiger}: 
\begin{equation}
    T_{2,ID}^{-1}=\frac{2\pi^{2}\mu_{0}h\eta\gamma^{2}}{9\sqrt{3}} *(\frac{3}{2})
\end{equation}
where $\eta$ represents the effective concentration of resonant spins contributing (in this case significantly lower than the total defect concentration), $\mu_{0}$ is the permeability of the free space, $h$ is Planck's constant, and $\gamma$ is the electron gyromagnetic ratio. As illustrated in Fig.~\ref{fig:wide-image}(c), the concentration of \vb\ spins is not a simple linear function of the ion fluence due to varying conversion efficiency, and so we used the PL intensity as a proxy for overall \vb\ concentration. Here we assume that for the 1e12 sample, each ion creates $\sim$ 2.5 vacancies on average based on SRIM calculations --- see SI). To estimate $\eta$, we then scaled this concentration by a factor of 0.18, calculated from the fraction of the total ODMR linewidth addressed by the microwave pulse excitation (see SI). The result is a quantitative estimate of the contribution of instantaneous diffusion to $T_2$, which we overlay on the measured values in Fig.~\ref{fig: spin}(c). Based on this model, we expect instantaneous diffusion to begin limiting $T_2$ for ion fluences above $10^{14}$~{ions/cm}$^2$, consistent with the measured values.
Advanced dynamical decoupling sequences, such as DROID, are required to effectively suppress such dipolar coupling interactions\cite{coherentspindynamicsnatureGong2022}. 

Having determined key optical and spin properties of the \vb\ as a function of ion fluence, we can use these values to understand how \vb\ implantation conditions can be optimised for magnetic field sensitivity. Previous studies\cite{magneticsensingvalueGottscholl2021,lowsensitivity,electronirradiationoptimization,magneticimagingactualHuang2022,senstivityoptimizationusinglaserandmicrowaveZhou2023,highsensitivityisotope} have demonstrated DC magnetic sensitivities ranging from 100~$\upmu$T/$\sqrt{\text{Hz}}$ to 3~$\upmu$T/$\sqrt{\text{Hz}}$ and predict an estimated sensitivity limit of around 20~nT/$\sqrt{\text{Hz}}$ under ideal conditions\cite{magneticsensingvalueGottscholl2021}.
We calculated the DC sensitivity using the following relation \cite{dcsensitivity} : 
\begin{equation}
    \eta_{DC} (\text{T}/\sqrt{\text{Hz}}) \approx \frac{h}{g \mu_B} \frac{1}{\sqrt{\mathcal{R}}\max \left| \frac{\partial \mathcal{I}}{\partial \nu_0} \right|}
\end{equation}
where g denotes the g-factor of the \vb\ centers ($\sim$ 2), $\mu_B$ is the Bohr magneton, $\mathcal{R}$ is the photon count rate and $\max \left| \frac{\partial \mathcal{I}}{\partial \nu_0} \right|$ represents the slope of the continuous wave ODMR signal (accounting for both contrast and linewidth). The estimated DC magnetic field sensitivity is shown in Fig.~\ref{fig: spin}(d)), indicating an optimal ion fluence between $10^{14}$ and $10^{15}$~{ions/cm}$^2$ and a DC sensitivity of around 30~$\upmu$T/$\sqrt{\rm Hz}$. This value reflects a trade-off between photon count rate, ODMR linewidth broadening, and contrast, all of which contribute to overall magnetic sensitivity performance. 
We also estimated AC magnetic sensitivity using the following relation\cite{highsensitivityisotope,Barry2020SensitivityMagnetometry}:
\begin{equation}
\eta_{\text{AC}}(\text{T}/\sqrt{\text{Hz}}) \approx \frac{\pi}{2 \gamma_e} \frac{1}{C_{\max} e^{-\left(\tau/T_2\right)} \sqrt{\mathcal{N}}} \frac{\sqrt{t_I + \tau + t_R}}{\tau},
\end{equation}
where $C_{\max}$ is the maximum measurement contrast, $\mathcal{N}$ is the average photon count collected,$\gamma_e$ is the electron gyromagnetic ratio, $\tau$ is the free evolution time for phase accumulation, optimally set near the coherence time $T_2$, and $t_I$ and $t_R$ are, respectively, the initialization and readout times. The optimal ion fluence was identified as $10^{14}$~{ions/cm}$^2$, leading to AC sensitivity of around 1~$\upmu\mathrm{T}/\sqrt{\mathrm{Hz}}$. 

Both AC and DC magnetic field sensitivity improve with increasing ion fluence, due to the increased \vb\ PL emission. In the case of AC sensing, an optimum is reached around $10^{14}$~{ions/cm}$^2$ which appears to be limited by a drop in coherence time caused by instantaneous diffusion --- i.e.\ by coupling to other, resonant, \vb\ spins. In contrast, for DC sensing the limiting factor as ion fluence increases appears to be related to lattice damage and its corresponding effect on the ODMR contrast. This raises the question of whether there is significant migration of \vb\ defects at the interface between implanted and unimplanted regions of the hBN flake. In principle, such migration could provide a route to decouple the creation of \vb\ spins from the corresponding lattice damage. We next explore such interfaces using PL to monitor \vb\ migration and Raman spectroscopy to provide a more general probe of changes in hBN lattice.



\begin{figure}[htbp] 
    \centering
    \includegraphics[width=0.4\textwidth]{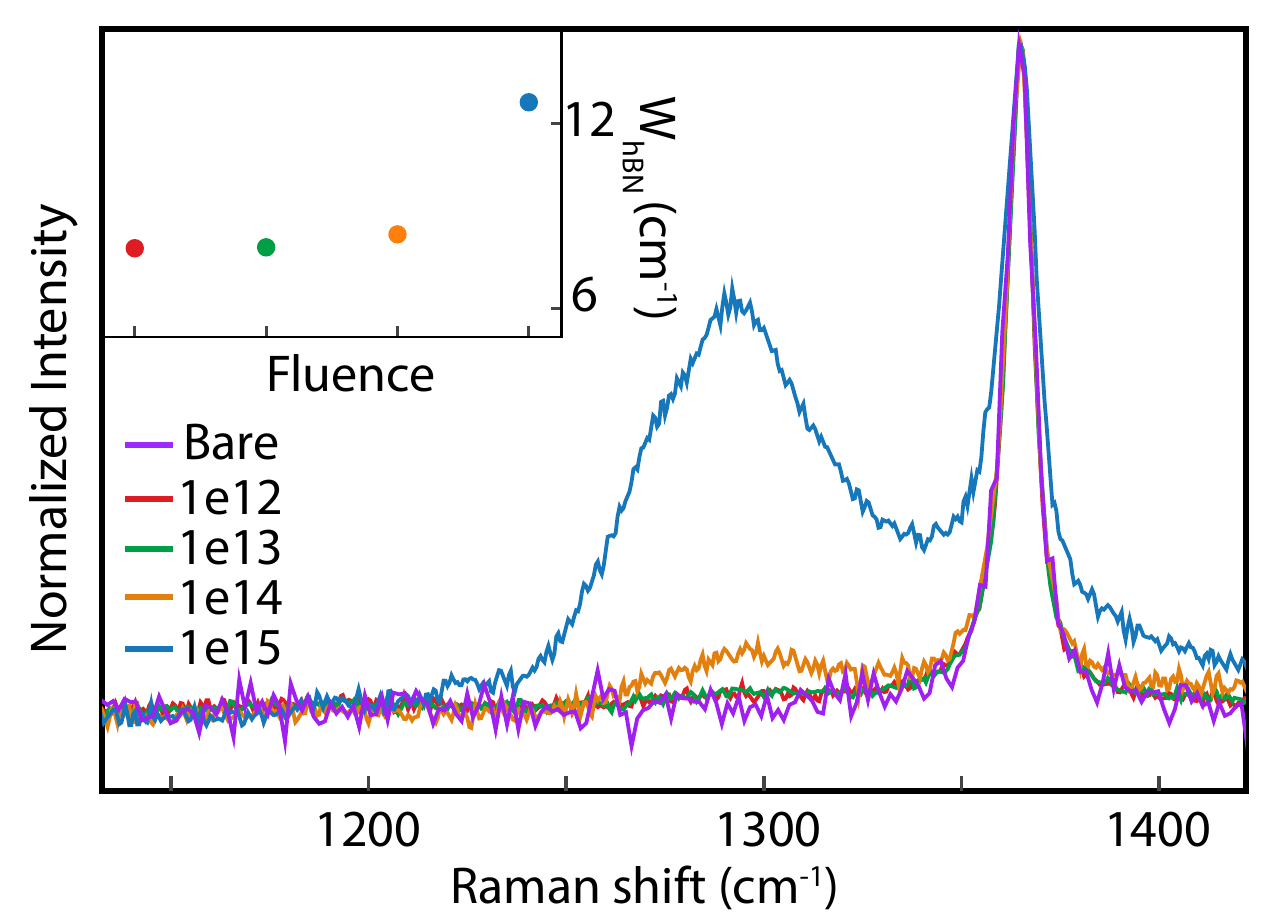} 
    \caption{Raman spectra for different ion-implanted hBN samples and one unimplanted sample. Inset shows the variation in the linewidth of the hBN peak (1365 cm$^{-1}$). 
    }
    \label{fig:raman}
\end{figure}

We first present Raman spectroscopy obtained well within the ion implanted region, as a function of ion fluence (see Fig.~\ref{fig:raman}). In unimplanted flakes, the characteristic hBN Raman peak ($E_{2g}$ mode) was observed at $\sim$1365~cm$^{-1}$, and we confirmed that any lattice strain induced by transferring the flake from a silicon substrate to the gold microstrip was sufficiently negligible that it had no impact on the Raman spectrum (see Fig.~SI7).
%
At an ion fluence of $10^{15}$~{ions/cm}$^2$, broadening of the hBN Raman peak is observed --- this can be attributed to increased lattice disorder and the formation of complex defect clusters\cite{raman2}, and is consistent with the reduction of ODMR contrast and $T_1$ for \vb\ spins presented above. At fluences of $10^{14}$~{ions/cm}$^2$ and above, an additional peak near $\sim$1295~cm$^{-1}$ also appears. This peak has been recently attributed to \vbn\ centres\cite{raman1,raman2} (see Fig.~SI8). 



\begin{figure}[t]
    \centering
    \includegraphics[width=0.49\textwidth]{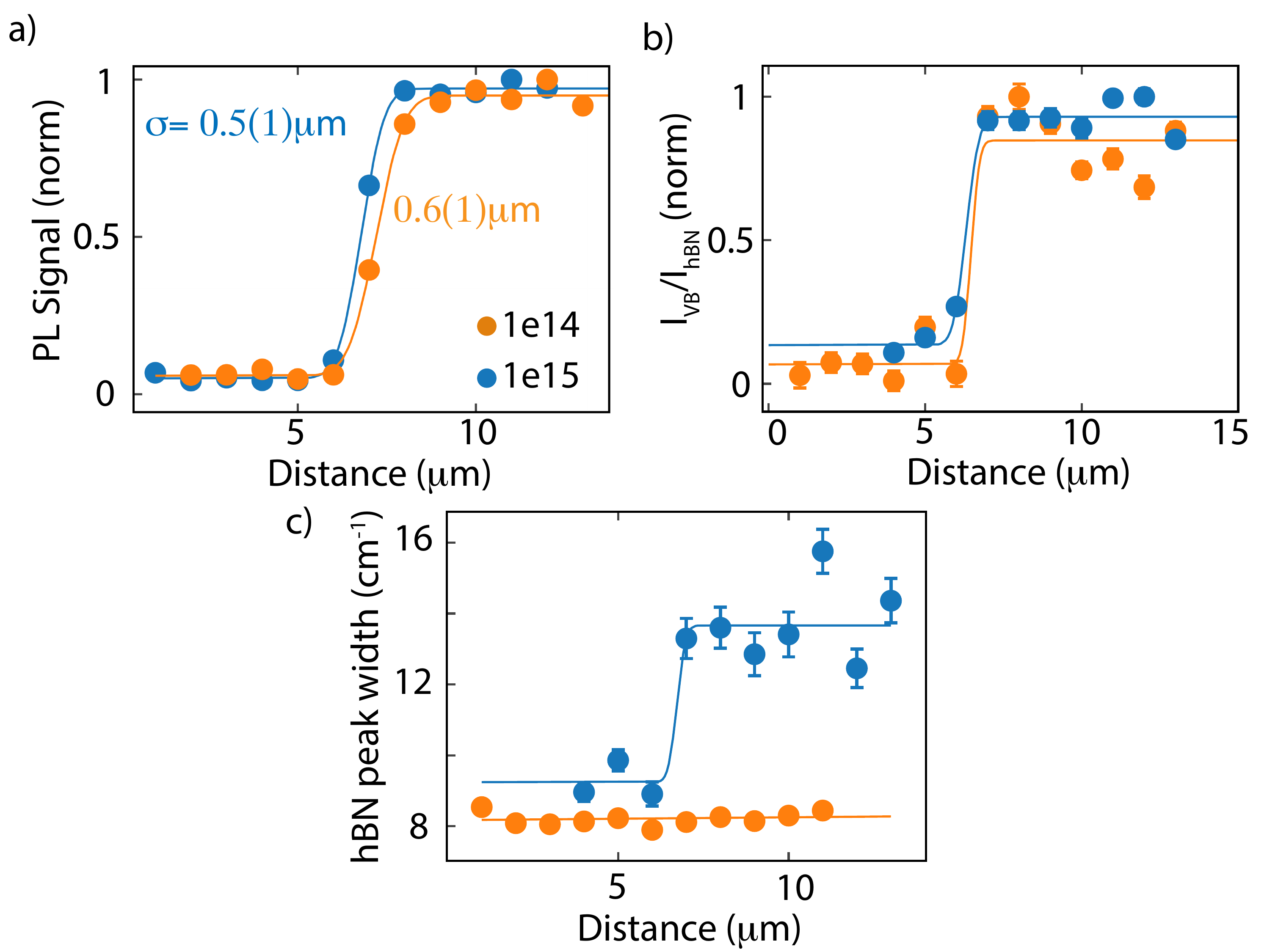} 
    \caption{PL and Raman studies around the boundary between implanted and pristine areas for samples 1e14 and 1e15, respectively.  a) Normalized PL signal from V$_{B}^{-}$ centres along a line. b) Raman peak intensity ratio I$_{V_{B}}$/I$_{hBN}$ (normalized w.r.t. maxima). c) The hBN peak width(FWHM) along the line. }
    \label{fig:migration}
\end{figure}

To investigate migration of defects during FIB ion implantation, we performed PL and Raman spectroscopy measurements along a line cut across the boundary between unimplanted and implanted regions, using a step size of $\sim$1 $\upmu$m. In Fig.~\ref{fig:migration} we show the measurements for samples 1e14 and 1e15, and fit the profiles using a Gaussian error function to obtain the width of the transition region (see SI).

The intensity of the PL signal for both the samples fell across the boundary to the unimplanted region, with a similar width of about $\sim$ 0.5~$\upmu$m (while $\sigma_{\text{laser}} \sim$ 0.3~$\upmu$m). This rules out significant migration of \vb\ centres toward the unimplanted lattice region under either of the implantation parameters studied. To analyse the spatial variation in the Raman spectra, we focus on two parameters: i) ratio of the intensity of the defect-related peak $I_{\rm V_B}$ and that of the primary hBN peak $I_{\rm hBN}$; and ii) the width of the hBN peak, which increases in the 1e15 sample (see inset of Fig.~\ref{fig:raman}). 
Each of these Raman features showed a similar step-like function with width $\sigma \sim 0.3~\upmu$m consistent with the beam width used, suggesting that the lattice damage from implantation is limited to the irradiated region for the range of fluences studied here (Fig.~\ref{fig:migration}(c)). . 

\section{Conclusion}
FIB implantation of He$^+$ ions into hBN is an effective way to pattern \vb\ centres which could act as quantum sensors embedded within 2D materials. We have explored how varying the ion fluence influences the properties of \vb\ centres relevant to their application as local magnetic sensors. We identify that with increasing concentration of \vb\ centres (through higher ion fluence), the residual defects begin to influence different parameters, ultimately reducing the measurement sensitivity. While higher ion fluence enhances the \vb\ centre-related brightness, it also leads to degradation of the spin parameters. The optimal ion implantation condition was determined to be $10^{14}$~{ions/cm}$^2$. Under these conditions, a DC field sensitivity of $30~\upmu \rm T/\sqrt{\rm Hz}$ was inferred, becoming limited by a reduction in ODMR spin contrast which we attributed to lattice damage induced by the ion implantation. In principle, this DC sensitivity could be improved by decoupling the creation of \vb\ centres from the formation of lattice defects, however, achieving this through the natural migration of \vb\ centres proved unsuccessful. A maximum AC sensitivity of $\sim 1~\upmu \rm T/\sqrt{\rm Hz}$ was inferred using an ion fluence of $10^{14}$~{ions/cm}$^2$. Here the sensitivity is again limited by a reduction in spin contrast, but now also by dipolar interactions between resonant \vb\ spins, posing a more fundamental barrier to further increases. Overall, the insights into these limiting factors in the sensitivity of ion-implanted \vb\ sensors will aid in the engineering of 2D magnetic sensor arrays with optimised magnetic sensing properties.


\begin{acknowledgments}
The authors would like to thank Patrick Hogan for helpful discussions and the cleanroom support staff at the London Centre for Nanotechnology for their technical assistance. This work was supported by the Engineering and Physical Sciences Research Council (EPSRC) through the Hub in Quantum Computing and Simulation (Grant No. EP/T001062/1) and through EP/T517793/1. S.D.\ thanks the Indian Institute of Technology, Madras, India, and the Science and Engineering Research Board (SERB Grant No. SRG/2023/000322), India, for start-up funding. S.D.\ and B.M.\ acknowledge the use of facilities supported by a grant from the Mphasis F1 Foundation given to the Center for Quantum Information, Communication, and Computing (CQuICC). K.W. and T.T. acknowledge support from the JSPS KAKENHI (Grant Numbers 21H05233 and 23H02052), the CREST (JPMJCR24A5), JST and World Premier International Research Center Initiative (WPI), MEXT, Japan.
\end{acknowledgments}

\section*{References}
\bibliography{main}

\end{document}


\maketitle

\section{Sample Fabrication}
Gold strip-lines were fabricated on bare SiO$_2$/ Si substrates using direct-write photolithography. 5nm of chrome was evaporated as an adhesion layer, followed by 210nm of gold by electron beam evaporation. Lift-off was performed in DMSO at 65$^o$C for 10 minutes. hBN flakes used in this study were grown under high pressure and high temperature \cite{taniguchi2007synthesis}. The flakes were first exfoliated on a bare SiO$_2$/ Si substrate and picked up by a PC stamp, aligned with an optical microscope and deposited on the gold strip-line \cite{purdie2018cleaning}. The PC stamp was melted onto the gold strip-line at 180$^o$C and residual PC was dissolved in chloroform. The hBN-gold strip-line structures were cleaned by acetone and IPA.

\begin{figure}[H]
    \centering
    \includegraphics[width=0.9\textwidth]{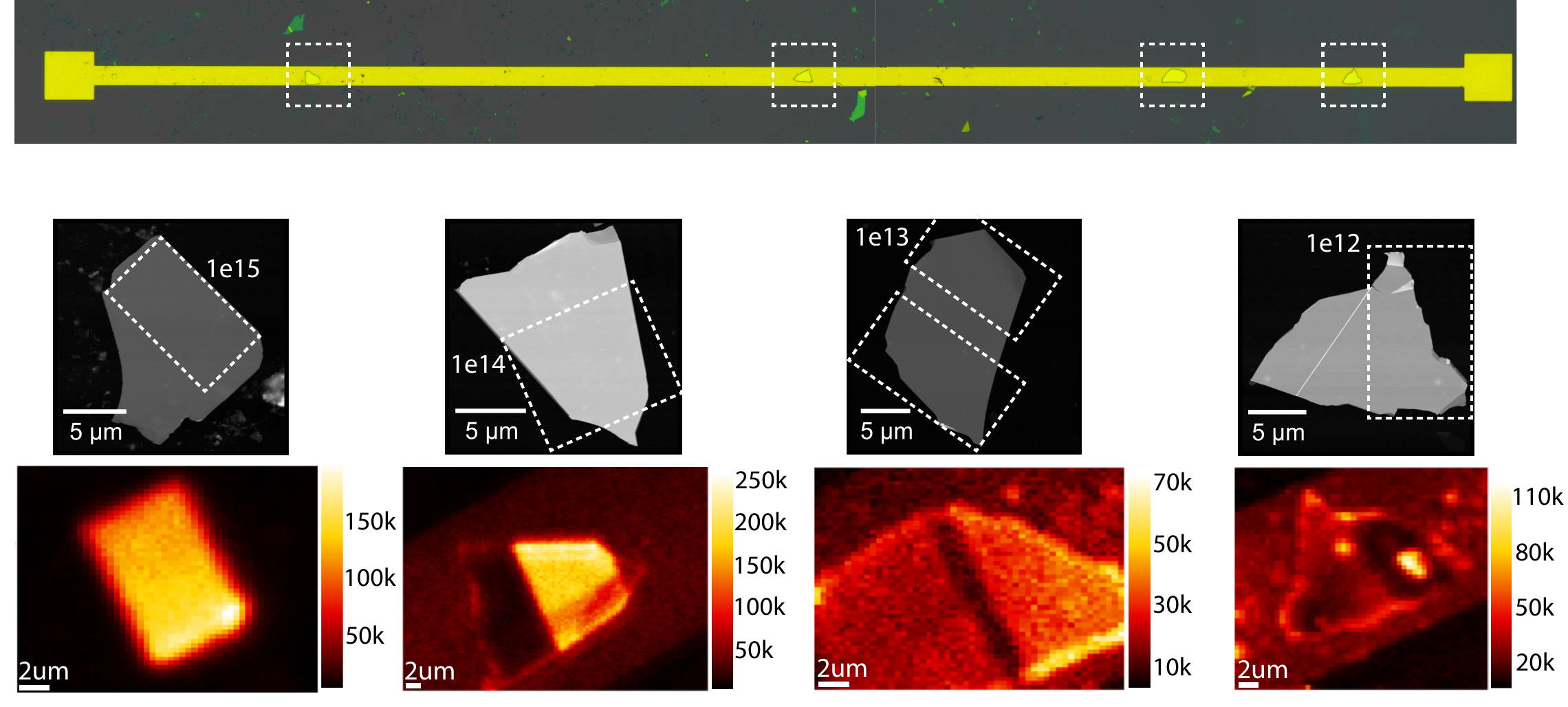} 
    \caption{ Optical Image of the gold microstrip (not to scale), displaying all flakes implanted with varying ion fluence. Below this, Atomic Force Microscopy (AFM) images and the corresponding confocal microscope images of the same flakes are shown. The dotted regions in the AFM images indicate the implanted area, which is also distinguishable in the confocal images through the change in photon counts.}
    \label{fig:sample}
\end{figure}

\begin{figure}[H]
    \centering
    \includegraphics[width=0.95\textwidth]{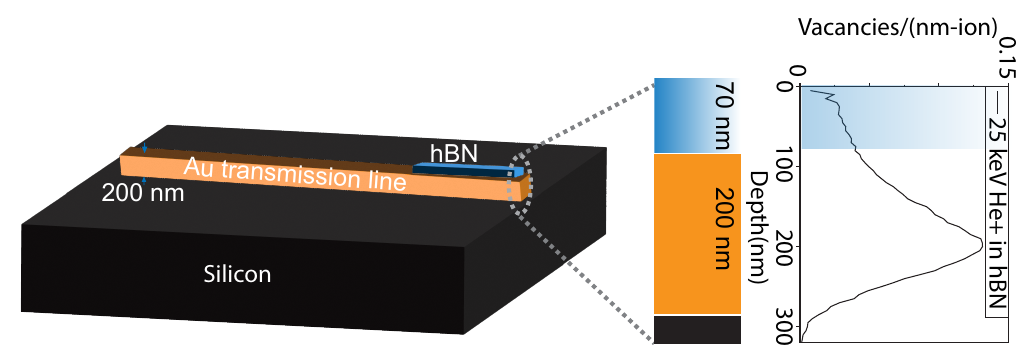} 
    \caption{ Cartoon representation of sample configuration. The hBN nanoflakes are placed on a $\sim$ 200~nm thick gold transmission line fabricated on the top of a silicon substrate. The vacancy profile, estimated using SRIM simulations, has also been shown.  }
    \label{SRIM}

\end{figure}

The concentration of boron vacancies was estimated using SRIM simulations, which revealed that each incident He$^+$ ion generates approximately 22 vacancies along its full penetration depth (maximum range $\sim$270~nm) in the hBN lattice. From the calculated vacancy profile (Fig.~\ref{SRIM}), we estimate that about 12\% of these vacancies are produced within the top 70~nm---the typical thickness of the hBN nanoflakes studied here---corresponding to $\sim$2.5 boron vacancies per implanted He$^+$ ion.

\begin{figure}[H]
    \centering
    \includegraphics[width=0.65\textwidth]{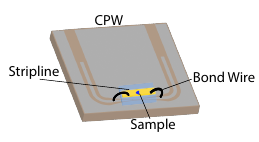} 
    \caption{The combined coplanar waveguide (CPW) microstrip design is used to apply microwaves for the manipulation of V$_{B}^{-}$ defects in these flakes. The CPW features a central constriction, and the gap between the conductor arms and is based on a previously developed design within the group.\cite{felixthesis}  }

\end{figure}

\section{Model System Hamiltonian}
We consider a model Hamiltonian describing a single $V_{B}^-$ defect (S = 1) interacting with the three first-shell nitrogen spin bath($^{14}$N) via hyperfine interactions. The Hamiltonian of the composite system can be written as:

\begin{equation}
    H = D\left(S_z^2 - \frac{2}{3}\right) + E \left(S_y^2 - S_x^2\right) + \gamma_e  \bm{B} \cdot \bm{S} + \sum_{i=1}^{3} \bm{S} \cdot \bm{A}^i \cdot \bm{I}^i +  \sum_{i=1}^{3} \gamma_{n}^i \bm{B} \cdot \bm{I}^i + \sum_{i=1}^{3} \bm{I}^i \cdot \bm{Q}^i \cdot \bm{I}^i.
\end{equation}

Here, $\bm{S}$ and $\bm{I}$ are the electron and nuclear spin operators, respectively. $D$ represents the zero-field splitting (ZFS) term, while $E$ accounts for the electric field susceptibility. $\gamma_e$ and $\gamma_n$ denote the electron and nuclear gyromagnetic factors, respectively. $\bm{A}$ is the hyperfine interaction tensor, $\bm{Q}$ is the quadrupole interaction tensor, and $\bm{B}$ represents the externally applied magnetic field (Zeeman interaction). The coupling parameters used in our numerical calculations are listed in Table~\ref{table:coupling_parameters}.
\begin{table*}[ht!]
    \centering
    \begin{tabular}{ |c|c|c|c|c|c|c|c|c| }
        \hline
        & $A_{xx}$ & $A_{yy}$ & $A_{zz}$ & $A_{xy}$ & $Q_{xx}$ & $Q_{yy}$ & $Q_{zz}$ & $Q_{xy}$ \\
        \hline
        $N^1$ & 46.944 & 90.025 & 48.158 & 0 & -0.46 & 0.98 & -0.52 & 0 \\
        \hline
        $N^2$ & 79.406 & 58.170 & 48.159 & -18.391 & 0.62 & -0.1 & -0.52 & -0.623 \\
        \hline
        $N^3$ & 79.406 & 58.170 & 48.159 & 18.391 & 0.62 & -0.1 & -0.52 & 0.623 \\
        \hline
    \end{tabular}
    \caption{Hyperfine and quadrupolar tensor elements used in the numerical calculations.\cite{theoreticalmodel1,theoreticalmodel2}}
    \label{table:coupling_parameters}
\end{table*}

\section{Optically Detected Magnetic Resonance (ODMR) Simulation}
We used QuTiP to compute the transition frequencies and probabilities based on Fermi's Golden Rule. Since the amplitude of transitions is directly proportional to the transition probability, we introduced an additional free parameter alongside $D$, $E$, and the linewidth. To simulate the optically detected magnetic resonance (ODMR) spectrum, we assumed all transitions follow a Lorentzian profile given by:

\begin{equation}
    L(\omega, A, \omega_0, \Delta\omega) = A \frac{(\Delta\omega/2)}{(\omega - \omega_0)^2 + (\Delta\omega/2)^2}
\end{equation}

where $\omega$ is the frequency, $A$ is the amplitude, $\omega_0$ is the central frequency, and $\Delta\omega$ is the full width at half maximum (FWHM). Based on this model, we used SciPy's curve-fitting functionality to fit the theoretical ODMR spectrum to experimental data, allowing us to extract the parameters $D$, $E$, and the linewidth.

\begin{figure}[H]
    \centering
    \includegraphics[width=0.45\textwidth]{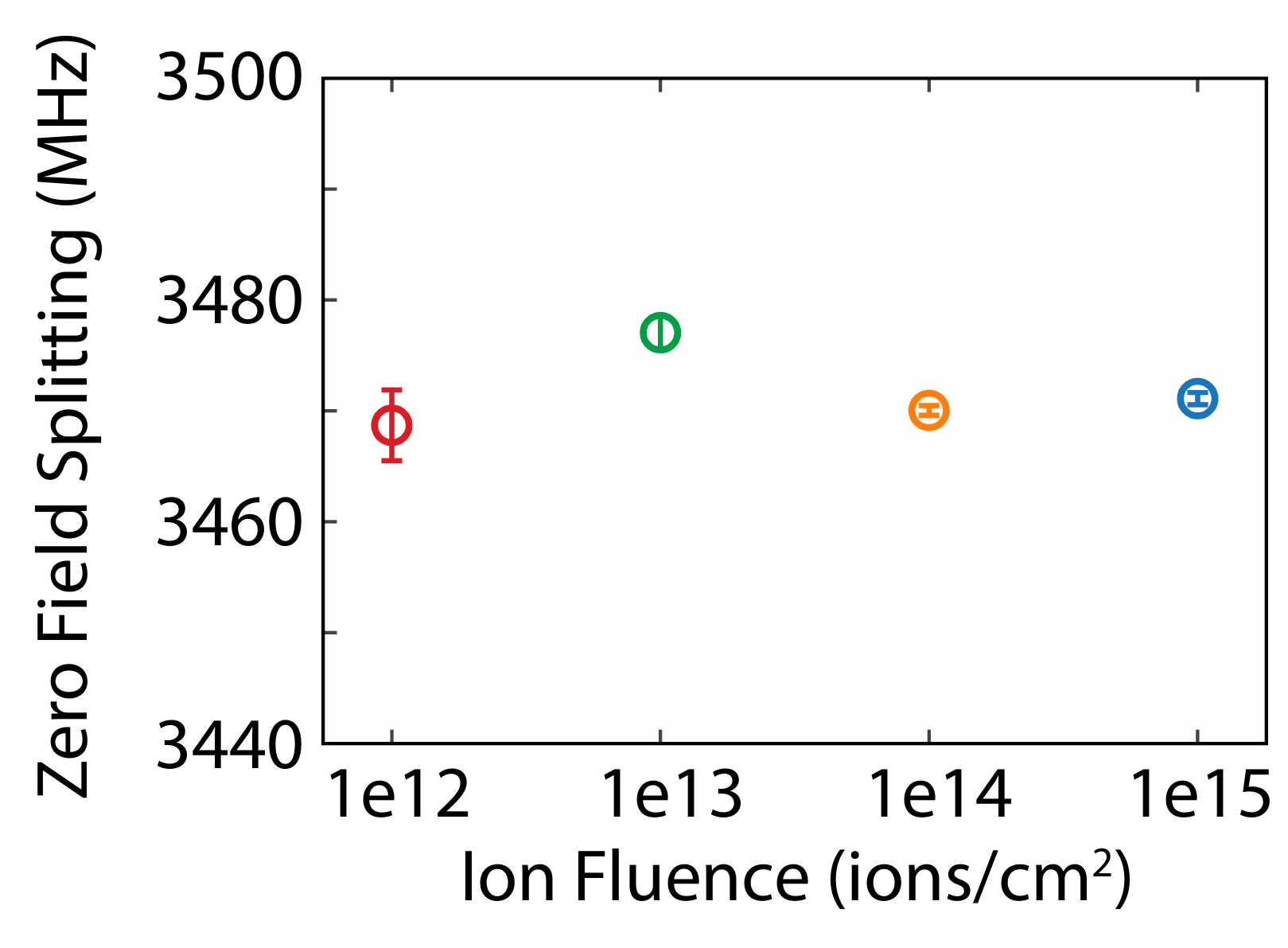} 
    \caption{Zero field splitting parameter(D) extracted from PODMR fitting, with varying ion fluence.}

\end{figure}

\section{Coherence time measurements}

\begin{figure}[H]
    \centering
    \includegraphics[width=0.95\textwidth]{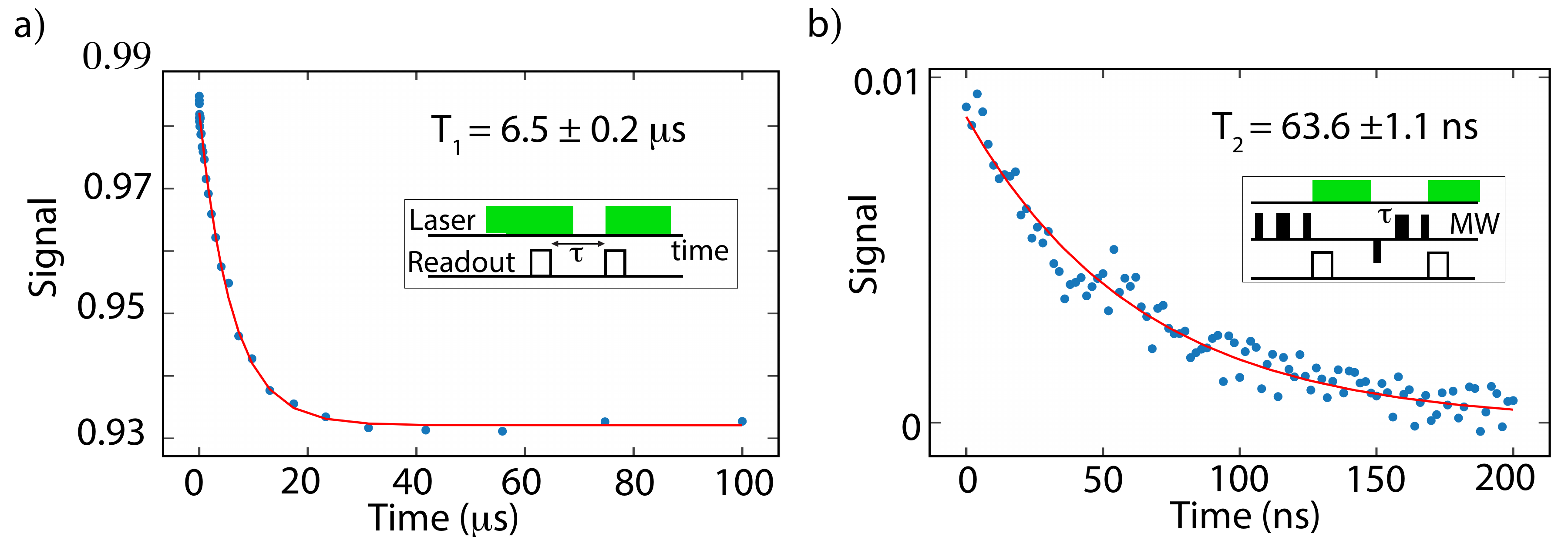} 
    \caption{Fitting of coherence time data for flake 1e15 with insets showing the pulse sequences used. a) T$_1$ data b) Hahn Echo T$_2$}

\end{figure}

\section{Instantaneous Spin Diffusion}

\begin{figure}[H]
    \centering
    \includegraphics[width=0.45\textwidth]{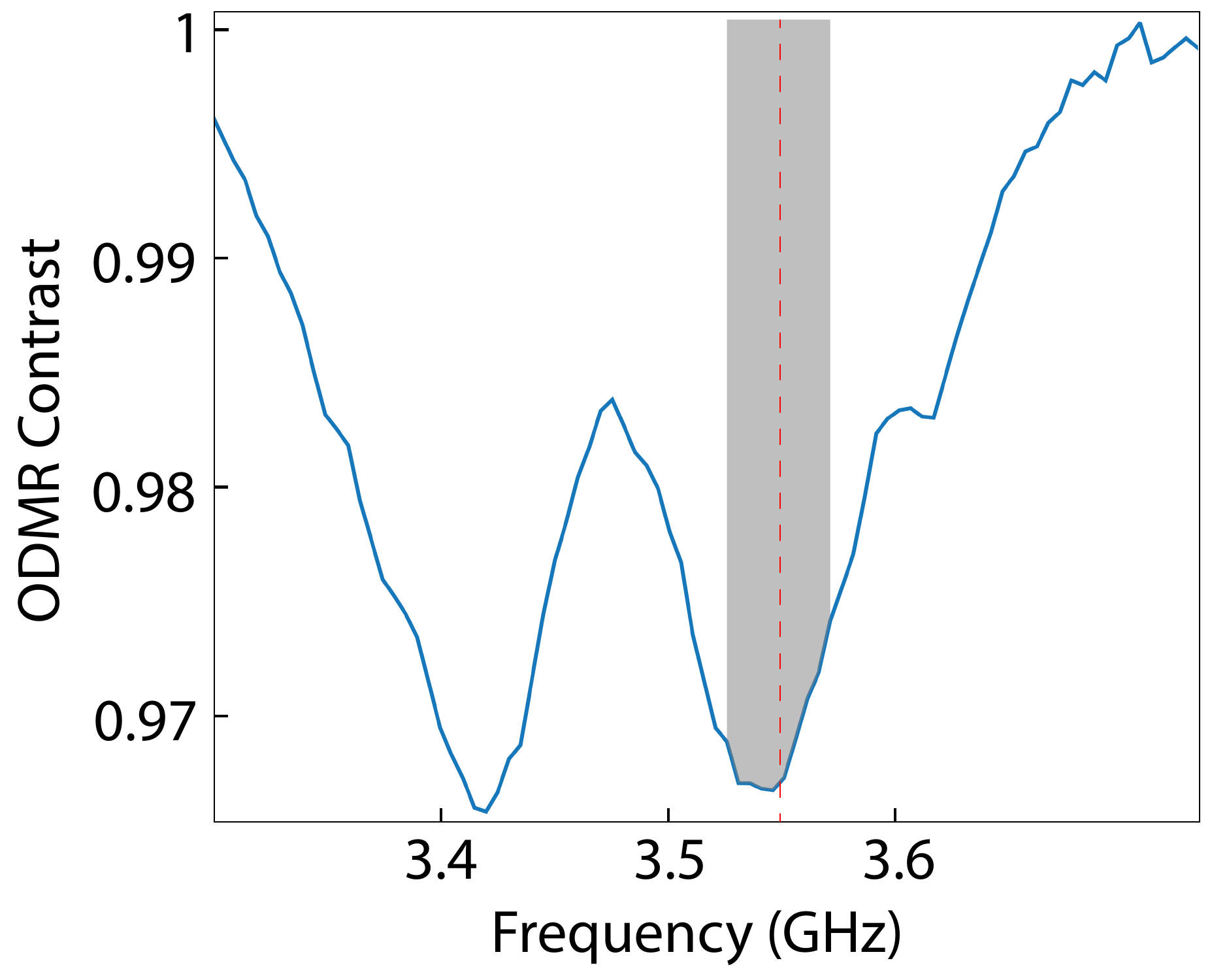} 
    \caption{PODMR data for sample 1e15, with the red line indicating the MW frequency used for the Hahn Echo experiment and the shaded grey region representing the calculated bandwidth of the pulse.}
\end{figure}

Based on area-under-the-curve calculations, we estimate that the Hahn echo $\pi$ pulse addresses approximately 18\% of the total spin ensemble. Interactions among these addressed spins are not refocused by the $\pi$ pulse and therefore contribute to instantaneous spin diffusion.

\section{Material characterization}

\begin{figure}[H]
    \centering
    \includegraphics[width=0.95\textwidth]{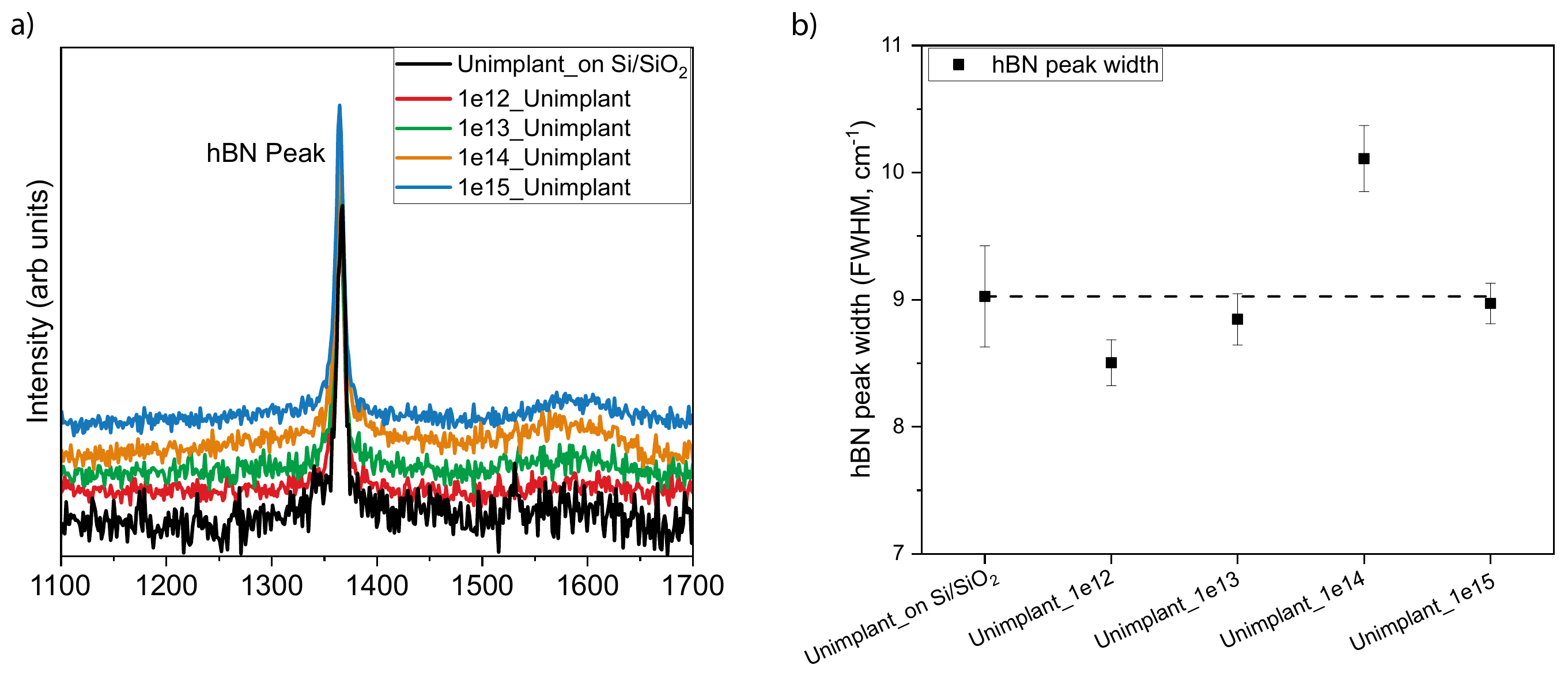} 
    \caption{(a) Raman spectra of different hBN nanoflakes (placed on gold transmission line) before FIB implantation. The Raman spectrum of an unimplanted hBN nanoflake, placed on a Silicon substrate (with a native thermal oxide layer), has been shown for comparison. (b) The hBN peak width for different unimplanted nanoflakes. }

\end{figure}

\begin{figure}[H]
    \centering
    \includegraphics[width=0.95\textwidth]{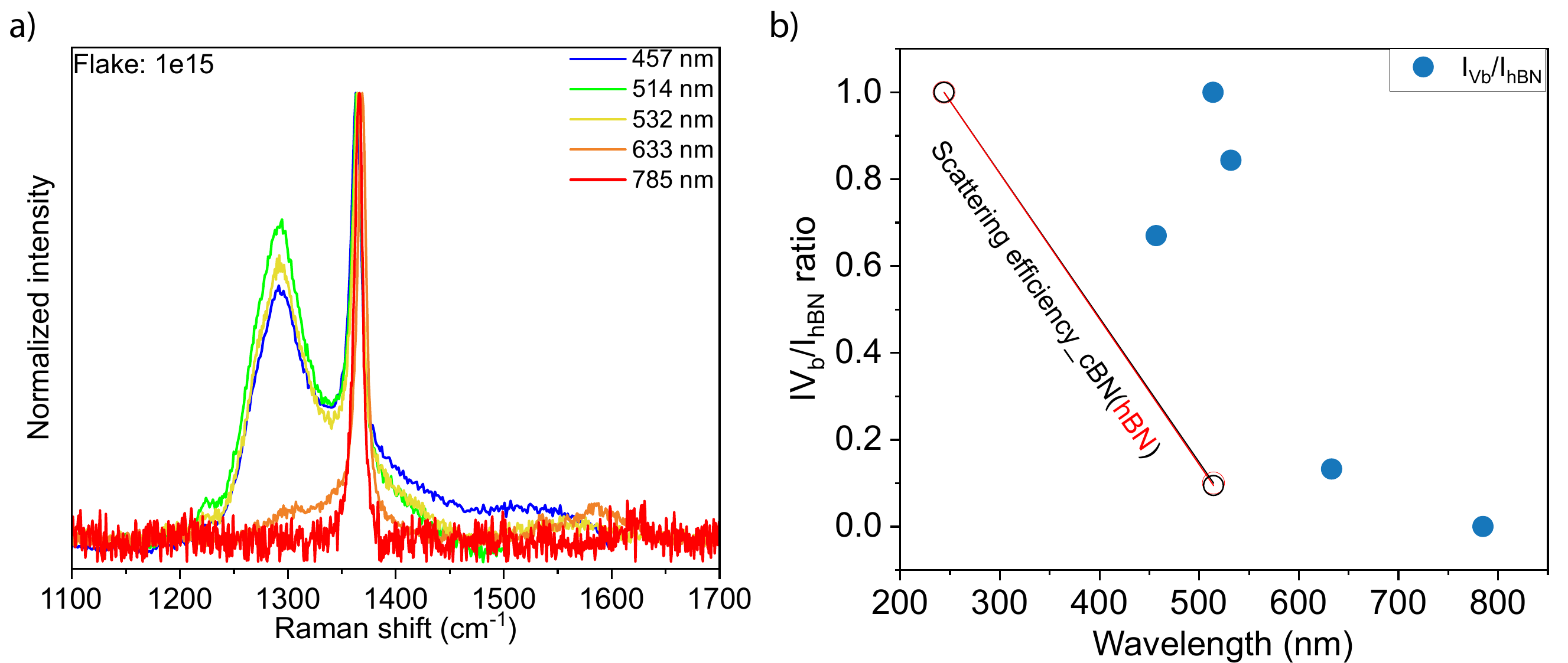} 
    \caption{Normalized Raman spectra under different laser excitation wavelengths between 457 - 785 nm. (b) The I$_{Vb}$/I$_{hBN}$ ratio for different laser excitation wavelengths. The maxima for I$_{Vb}$/I$_{hBN}$ was found to be at 514 nm. To compare the trend with wavelength dependent Raman signal intensity of cBN and hBN, we have compared the Raman scattering efficiency of both phases (shown as the black and red dotted lines). The scattering efficiency is taken as the proxy of Raman signal intensity for comparison.\cite{reich2005resonant}}
    \label{scattering}

\end{figure}

To further investigate the origin of the peak $\sim$ 1295~cm$^{-1}$, multiwavelength Raman spectroscopy was performed using laser wavelengths from 457~nm to 785~nm (see Fig.\ref{scattering} ). The peak intensity increases with excitation wavelength up to $\sim$500~nm and decreases thereafter, suggesting a resonance maximum around 500~nm, consistent with previous findings \cite{raman2}. This trend contrasts with typical hBN or cBN Raman behavior, which peaks in the UV regime due to their wide bandgap ($\sim$200~nm)\cite{ramanuv}. 

\section{Migration Analysis}
We modeled the expected distribution of defects under Gaussian beam illumination using the following function:

\begin{equation}
f(x) = \frac{1}{2} \left[ 1 + \text{erf}\!\left( \frac{x - x_0}{\sqrt{2}\sigma} \right) \right] \cdot h + C_0,
\end{equation}

where $x_0$ is the step offset, $\sigma$ is the standard deviation of the Gaussian profile, $h$ is the step height, and $C_0$ is the baseline offset.

The width of the underlying defect distribution can then be extracted from the measured profile using

\begin{equation}
\sigma_{\text{distribution}} = \sqrt{\sigma_{\text{measured}}^2 - \sigma_{\text{laser}}^2},
\end{equation}

where $\sigma_{\text{measured}}$ is the standard deviation obtained from fitting the Gaussian error function, and $\sigma_{\text{laser}}$ is the standard deviation of the laser’s Gaussian beam profile.

\bibliographystyle{plain}
\bibliography{supplementary}
